\documentclass[12pt]{article}

\usepackage{amssymb}

\begin{document}

\begin{titlepage}

\begin{flushright}
arXiv:1108:3827
\end{flushright}
\vskip 2.5cm

\begin{center}
{\Large \bf Bounding Lorentz Violation at Particle Colliders\\
By Tracking the Motion of Charged Particles}
\end{center}

\vspace{1ex}

\begin{center}
{\large Brett Altschul\footnote{{\tt baltschu@physics.sc.edu}}}

\vspace{5mm}
{\sl Department of Physics and Astronomy} \\
{\sl University of South Carolina} \\
{\sl Columbia, SC 29208} \\
\end{center}

\vspace{2.5ex}

\medskip

\centerline {\bf Abstract}

\bigskip

In the presence of Lorentz violation, the motion of a charged particle in a magnetic
field is distorted. By measuring the eccentricities of particles' elliptical orbits
and studying how those eccentricities vary with the absolute orientation of the
laboratory, it is possible to constrain the Lorentz-violating $c_{JK}$ parameters.
For each observed species, this method can provide constraints on four linear
combinations of coefficients for which, in some species, there are presently no
two-sided bounds.

\bigskip

\end{titlepage}

\newpage

\section{Introduction}

Recently, there has been a surge of interest in the possibility that some
symmetries (such as Lorentz and CPT invariances), while apparently
fundamental, might not actually hold exactly. Despite a large number of experimental
searches, no compelling evidence for violation of
Lorentz invariance or CPT has been found. However, new Lorentz and CPT tests are still
underway, and the study of these symmetries remains an active area of research---both
experimentally and theoretically.

If violations of Lorentz invariance or CPT are uncovered, that would be a discovery of
the utmost importance. Such symmetry violations would
be immediate evidence for new physics beyond the standard model with a completely
new structure. The mere existence of such symmetry breaking would tell us a tremendous
amount about the character of fundamental physics. For example,
a breaking of Lorentz symmetry could be tied to the structure of
quantum gravity.

Unlike some possible forms of exotic new physics that have been suggested, Lorentz
and CPT violations can be studied in the framework of quantum field theory.
Violations of these symmetries may be described in an effective field theory called
the standard model extension (SME). The SME contains Lorentz- and CPT-violating
corrections to the standard model, parameterized by small tensor-valued
background fields~\cite{ref-kost1,ref-kost2}. This field theory framework can
also be expanded to cover gravity~\cite{ref-kost12}.
Both the renormalizability~\cite{ref-kost4,ref-colladay2} and
stability~\cite{ref-kost3} of the SME have been studied.
The minimal SME (which contains only gauge-invariant, renormalizable forms of Lorentz
violation) has become the standard framework used for parameterizing the results of
experimental Lorentz tests.

Recent searches for Lorentz violation have included studies of matter-antimatter
asymmetries for trapped charged
particles~\cite{ref-bluhm1,ref-gabirelse,ref-dehmelt1} and bound state
systems~\cite{ref-bluhm3,ref-phillips},
measurements of muon properties~\cite{ref-kost8,ref-hughes}, analyses of
the behavior of spin-polarized matter~\cite{ref-heckel3},
frequency standard comparisons~\cite{ref-berglund,ref-kost6,ref-bear,ref-wolf},
Michelson-Morley experiments with cryogenic
resonators~\cite{ref-muller3,ref-herrmann2,ref-herrmann3,ref-eisele,ref-muller2},
Doppler effect measurements~\cite{ref-saathoff,ref-lane1},
measurements of neutral meson
oscillations~\cite{ref-kost10,ref-kost7,ref-hsiung,ref-abe,
ref-link,ref-aubert}, polarization measurements on the light from cosmological
sources~\cite{ref-carroll2,ref-kost11,ref-kost21,ref-kost22},
high-energy astrophysical
tests~\cite{ref-stecker,ref-jacobson1,ref-altschul6,ref-altschul7,ref-klinkhamer2},
precision tests of gravity~\cite{ref-battat,ref-muller4}, and others.
The results of these experiments set constraints on the various SME coefficients, and
up-to-date information about
most of these constraints may be found in~\cite{ref-tables}.

There are many precise bounds on the forms of Lorentz violation that affect stable
particles, but Lorentz violation with unstable particles is much more difficult
to study. There have been some impressive measurements made using muon spin
precession, neutral meson oscillations, pion decay~\cite{ref-altschul16},
and neutrons (both free neutrons~\cite{ref-altarev} and bound ones in stable
nuclei). However, there are still relatively few constraints
on the SME parameters for most second- and third-generation fields. Unstable
particles are produced and detected in copious numbers in collider experiments, and
so it is natural to look for ways to test Lorentz symmetry using new or existing
collider data sets. (Data from accelerators has already been used to place bounds
on electron and photon Lorentz
violations~\cite{ref-hohensee1,ref-altschul20,ref-bocquet}.)

Collider data that is considered to have little value in searches for novel (but still
Lorentz-invariant) physics may actually be fairly sensitive to violations of
Lorentz symmetry. In the paper, we shall identify an observable that may be used to
test Lorentz invariance for unstable charged species.
In magnetized detectors, ultrarelativistic particles trace out
synchrotron orbits, and the shapes of these orbits are affected by the SME
coefficients. We shall show how an analysis of this effect can be
used to place new constraints on SME parameters.

Section~\ref{sec-orbit} of this paper will introduce the form of Lorentz violation
that is most relevant for particle tracking experiments. The shapes of
synchrotron orbits in the presence of this kind of Lorentz violation have already
been calculated~\cite{ref-altschul5}, and it shall be shown that the modified motion
gives rise to orbits with nonvanishing eccentricities. The analysis of these
orbits
continues in section~\ref{sec-obs}, which describes how sidereal variations in
orbital eccentricities can be related to the underlying SME parameters. This
section is subdivided into two cases:  section~\ref{sec-LVdom} considers the case
in which Lorentz violation is the dominant source of orbital deformation; and
section~\ref{sec-systdom} deals with the possibility that other systematic
effects may largely determine an orbit's eccentricity, with Lorentz violation as an
additional small perturbation. Some comments and conclusions are given in
section~\ref{sec-concl}.

\section{Lorentz-Violating Synchrotron Motion}
\label{sec-orbit}

We shall be studying the quantum electrodynamics (QED) sector of the
SME. The QED Lagrange density containing the form of Lorentz violation that is most
relevant at high energies is
\begin{equation}
\label{eq-L}
{\cal L}=-\frac{1}{4}F^{\mu\nu}F_{\mu\nu}+
\bar{\psi}[(\gamma^{\mu}+c^{\nu\mu}\gamma_{\nu})(i\partial_{\mu}-qA_{\mu})-m]\psi.
\end{equation}
Spin-independent Lorentz violation at high energies is controlled primarily by the
$c^{\nu\mu}$ coefficients. There is a separate set of these coefficients for each
fermion species. The importance of the $c$ coefficients grows at high
energies. There is another, similar set of coefficients---the $d^{\nu\mu}$---which
also become more important at high energies, but the effects of the $d$ parameters
depend on helicity and average to zero for unpolarized particles. There is also the
potential for Lorentz violation in the electromagnetic sector; however, photon
Lorentz violation can be
constrained sufficiently accurately in low-energy, high precision experiments
that it may be neglected in this analysis of accelerator-based Lorentz tests.

At leading order, $c^{\nu\mu}$ may be taken to be traceless and symmetric, so it
contains nine independent parameters for each species. The energy of a particle,
including the leading order effects of $c$ is
\begin{equation}
E=\sqrt{(m^{2}+\vec{\pi}^{2})(1-2c_{00})-2c_{jk}\pi_{j}\pi_{k}}-2c_{0j}\pi_{j},
\end{equation}
where $\vec{\pi}$ is the mechanical momentum $\vec{p}-q\vec{A}$.
The corresponding velocity is~\cite{ref-altschul4}
\begin{equation}
v_{j}=
\frac{\pi_{j}}{\sqrt{m^{2}+\vec{\pi}^{2}}}-c_{00}\frac{\pi_{j}}{\sqrt{m^{2}+
\vec{\pi}^{2}}}-2c_{jk}\frac{\pi_{k}}{\sqrt{m^{2}+\vec{\pi}^{2}}}+c_{kl}
\frac{\pi_{j}\pi_{k}\pi_{l}}{(m^{2}+\vec{\pi}^{2})^{3/2}}-2c_{0j}.
\end{equation}
For stable charged particles, there are many possible ways to constrain the $c$
coefficients. However, for unstable particles, there are very few bounds; in most
cases, essentially the only bounds come from the absence of the photon decay
process $\gamma\rightarrow X^{+}+X^{-}$, which may be allowed for initial photon
energies $E\gtrsim m_{X}/\sqrt{|c_{X}|}$. Here $c_{X}$ represents a linear
combination of the $c$ coefficients for the particle species $X$; precisely
which linear combination can be constrained with a given observation depends on the
direction of the photons involved, although the coefficients are generally of
${\cal O}(1)$.
However, the photon decay bounds get worse for heavier particles,
since the threshold energy increases with the particle mass $m_{X}$; bounds based on
the survival of 10--100 TeV photons over astrophysical distances are typically
at a level $\sim 10^{-9}\,[m_{X}/(1\,{\rm GeV})]^{2}$. Moreover, these
are one-sided bounds, and a sufficiently large value of $c_{00}$
(the coefficient of pure
boost invariance violation) could hide the effects of any other nonzero $c$
coefficients.

This means that any two-sided constraints on the $c$ coefficients for unstable
species will exclude new regions of the SME parameter space. Constraints
comparable to (or stronger than) the photon survival bounds would be especially
interesting. For muons, achieving this degree of precision may be a challenge,
but for heavier species, the
astrophysical bounds grow weaker quite rapidly. Accelerator constraints may be
competitive, since they would not be
expected to depend so strongly on the particle mass.

In this paper, we shall be concerned with the $c_{jk}$ coefficients
only---specifically, the five linear combinations of these coefficients that
break rotational isotropy. The orbits of relativistic charged
particles in a magnetic field are not as sensitive to other forms of Lorentz
violation. For example, the $d$ coefficients have no net affect on
particle orbits when there are equal populations of positive and negative helicity
fermions. The parameter $c_{00}$ breaks boost invariance only, and so it cannot be
constrained be measurements of spatial isotropy. Moreover, the spatial trace $c_{jj}$
is equivalent to $c_{00}$ by the (four-dimensional) tracelessness of $c^{\nu\mu}$.

The trickiest parameters to rule out of consideration
are the boost anisotropy parameters
$c_{0j}$. These are odd under parity (P) and time reversal (T),
and they do not affect the ellipticities of complete orbits. Since $c_{0j}$ provides
merely a constant offset to the
velocity of a particle, at nonrelativistic energies where no particles are created or
destroyed, $c_{0j}$ is completely unobservable. This is a consequence of
the Galilean invariance of nonrelativistic physics.
While $c_{0j}$ does have physical effects upon relativistic experiments, these coefficients will still only affect parity-odd observables.

In fact, if energy losses are neglected, the effect of the $c_{0j}$ coefficients
alone is to shift the velocity to $v_{j}=\pi_{j}/\sqrt{m^{2}+\vec{\pi}^{2}}-c_{0j}$.
The equation of motion
in a magnetic field, $\frac{d\vec{\pi}}{dt}=q(\vec{v}\times\vec{B})$ is not changed
by the
Lorentz violation (as a consequence of gauge invariance). So $\frac{d\vec{v}}{dt}=
\frac{q}{\gamma m}(\vec{v}\times\vec{B})$ continues to hold, even in the presence of
the $c_{0j}$. In other words, when synchrotron losses are neglected, $c_{0j}$ has no
effect on a particle's orbit.
If energy losses are included,
$\frac{dv_{j}}{dt}$ does acquire a new term $(\dot{E}/E)c_{0j}$,
in addition to the terms ordinarily present. This is potentially observable, but its
effects are suppressed by the energy loss rate, and the novel force does not affect the eccentricity of adiabatically shrinking orbits.

We shall therefore neglect the $c_{00}$ and $c_{0j}$, restricting our attention to
$c_{jk}$ Lorentz violation only.
The motion of a particle with this kind of Lorentz violation in a constant
background magnetic field $\vec{B}$ can be calculated exactly, if the
radiation by the accelerated charges is neglected. To first order in the $c_{jk}$
coefficients, the velocity of an orbiting charge is~\cite{ref-altschul5}
\begin{equation}
\left[
\begin{array}{c}
v_{1}(t) \\
v_{2}(t) \\
v_{3}(t)
\end{array}
\right]=\left[
\begin{array}{c}
v_{10}\left(\cos\omega t+2c_{12}\sin\omega t\right) \\
-v_{10}(1+c_{11}-c_{22})\sin\omega t \\
2v_{10}\left[c_{23}\sin\omega t-c_{13}\left(\cos\omega t-1\right)\right]+v_{30}
\end{array}
\right].
\label{eq-v}
\end{equation}
The direction $\hat{B}$ of the magnetic field is taken to be along the $z$-axis, and
the initial conditions are that $\vec{v}(0)=(v_{10},0,v_{30})$.
The frequency
$\omega=(1-c_{11}-c_{22})(eB/E)$
differs from the usual expression by small Lorentz-violating corrections, although
the energy $E=\sqrt{m^{2}+(\delta_{jk}-2c_{jk})\pi_{j}\pi_{k}}$ remains a constant of
the motion.

As in conventional
synchrotron motion, there are two components to the charged particle's
trajectory---periodic motion superposed with a constant velocity along
the direction $\hat{B}$ of the field. The periodic elliptical motion is not
precisely restricted to the plane perpendicular to $\vec{B}$, but the tilt
out of this plane is small. The projection of the orbit in the
plane normal to $\vec{B}$ is an ellipse, which generally has nonzero
eccentricity; this eccentricity is the key indicator that rotation
symmetry does not hold exactly. When radiation emission is included, the
size of an elliptical track will decrease adiabatically if the energy
losses are gradual enough. The ellipse will shrink slowly, but its
eccentricity and orientation will remain unchanged.

The shape of the orbit in the directions perpendicular to
$\vec{B}$ is governed by the projection of $c_{jk}$ onto those two directions:
\begin{equation}
c^{\perp}_{jk}=c_{jk}-2(c_{jl}\hat{B}_{l})\hat{B}_{k}+(c_{lm}\hat{B}_{l}\hat{B}_{m})
\hat{B}_{j}\hat{B}_{k}.
\end{equation}
It is evident from setting $c_{12}=0$ in (\ref{eq-v}) that the
charge's oscillatory motions
along the the eigenvectors of $c^{\perp}$ (neglecting the eigenvector $\hat{B}$ with
vanishing eigenvalue) are out of phase by $\frac{\pi}{2}$. In fact,
when the orbital ellipse
is projected into the plane normal to $\hat{B}$, the major and minor axes are
oriented along those eigenvectors of
$c^{\perp}$. Calling unit vectors in these directions $\hat{a}$ for the major
axis and $\hat{b}$ for the minor, the ratio of the two axes is
\begin{equation}
\frac{a}{b}=1-c^{\perp}_{jk}\hat{a}_{j}\hat{a}_{k}
+c^{\perp}_{jk}\hat{b}_{j}\hat{b}_{k}.
\end{equation}
$c^{\perp}_{jk}\hat{a}_{j}\hat{a}_{k}$ and $c^{\perp}_{jk}\hat{b}_{j}\hat{b}_{k}$
are the two (potentially nonzero) eigenvalues of $c^{\perp}$.
The eigenvector of $c^{\perp}$ corresponding to the smaller
eigenvalue indicates the major axis direction; the larger eigenvalue indicates the
minor axis.

So the orbital eccentricity squared is
$\epsilon^{2}=2|\lambda_{1}-\lambda_{2}|$---proportional to the
difference between the larger and smaller eigenvalues of the $2\times2$
matrix formed by projecting $c_{jk}$ into the plane normal to $\vec{B}$. Henceforth,
we shall refer to this $2\times2$ matrix as $c^{\perp}$, excising the row and column
of zeros the correspond to the $\hat{B}$-direction. The matrix may
be constructed explicitly by choosing two unit vector $\hat{e}_{1}$ and $\hat{e}_{2}$,
so that $\hat{e}_{1}$, $\hat{e}_{2}$, and $\hat{B}$ form an orthonormal
basis. Then the $2\times2$ projection of $c_{jk}$ is
\begin{equation}
c^{\perp}=\left[
\begin{array}{cc}
c_{jk}\hat{e}_{1j}\hat{e}_{1k} & c_{jk}\hat{e}_{1j}\hat{e}_{2k} \\
c_{jk}\hat{e}_{1j}\hat{e}_{2k} & c_{jk}\hat{e}_{2j}\hat{e}_{2k}
\end{array}
\right].
\end{equation}
The eigenvalue difference is then
$(\lambda_{1}-\lambda_{2})=\sqrt{(c^{\perp}_{11}-c^{\perp}_{22})^{2}+4
(c^{\perp}_{12})^{2}}$ in terms of the elements of $c^{\perp}$.

\section{Experimental Observables}
\label{sec-obs}

The eccentricity of a synchrotron orbit is an experimental observable, which can be
related to the underlying $c_{jk}$ coefficients for the revolving species. However,
the relationship between $\epsilon$ and the $c_{jk}$ in section~\ref{sec-orbit} was
derived under the assumption that Lorentz violation was the largest effect driving
the orbits to have systematic ellipticities. The experimental observables in such a
regime are derived in section~\ref{sec-LVdom}. However, if there are systematic
effects that dominate the orbital eccentricity, the observables are different and
somewhat simpler; they are explained in section~\ref{sec-systdom}. Which of the
two eccentricity regimes is physically relevant must be inferred from observational
data.

The same subset of the $c_{jk}$ coefficients can be constrained in either case.
Experimental constraints on the SME parameters are conventionally expressed in a system of Sun-centered celestial equatorial coordinates~\cite{ref-bluhm4}.
Coordinates
in this system are denoted by capital letters. From the origin at the center of the
Sun, the $Z$-axis points along the Earth's rotation axis; the $X$-axis
points toward the vernal equinox point on the celestial sphere; and the
$Y$-axis is determined by the right hand rule.

\subsection{Eccentricities Dominated by Lorentz Violation}
\label{sec-LVdom}

If Lorentz violation is the largest source of orbital eccentricity,
$\epsilon$ can be expressed in terms of the $c_{JK}$ parameters in
the Sun-centered frame and the components of $\hat{B}$.
The explicit form of the eigenvalue difference squared,
$(\lambda_{1}-\lambda_{2})^{2}$, is
\begin{eqnarray}
(\lambda_{1}-\lambda_{2})^{2} & = & \sum_{J=1}^{3}(1-\hat{B}_{J}^{2})^{2}c_{JJ}^{2}+
\sum_{J\neq K}\left\{-4\hat{B}_{J}\hat{B}_{K}(1-\hat{B}_{J}^{2})c_{JJ}c_{JK}
\right. \nonumber\\
& & \left.
+[2\hat{B}_{J}^{2}\hat{B}_{K}^{2}-(1-\hat{B}_{J}^{2})(1-\hat{B}_{K}^{2})]c_{JJ}c_{KK}
+2(1-\hat{B}_{J}^{2})(1-\hat{B}_{K}^{2})c_{JK}^{2}\right\} \nonumber\\
\label{eq-coeffs}
& & +
\sum_{J\neq K\neq L\neq J}\left[4\hat{B}_{K}\hat{B}_{L}(1+\hat{B}_{J}^{2})c_{JJ}c_{KL}
-4\hat{B}_{J}\hat{B}_{K}(1-\hat{B}_{L}^{2})c_{JL}c_{KL}\right].
\end{eqnarray}
The sums run over all combinations of distinct indices, and there are no additional
sums implied by repeated indices. Note that this means that each of the
$c_{JJ}c_{KK}$, $c_{JK}^{2}$, and $c_{JL}c_{KL}$ terms appears twice, with $J$ and $K$
exchanged; the total coefficient for a given one of these terms is thus twice what
what might be naively assumed from a casual look at (\ref{eq-coeffs}).

At any given instant, the magnetic field provides a fixed axis in space
$\vec{B}=B\hat{B}$. This axis rotates with the Earth. If $\vec{B}$ lies
in the plane tangent to the Earth's surface at the location of the lab (at
colatitude $\chi$), then the time-dependent unit vector $\hat{B}(t)$ is
\begin{eqnarray}
\hat{B}(t) & = & -(\cos\psi\cos\Omega t+\sin\psi\cos\chi\sin\Omega t)\hat{X}+
\nonumber\\
& & (\cos\psi\sin\Omega t-\sin\psi\cos\chi\cos\Omega t)\hat{Y}+
\sin\psi\sin\chi\hat{Z} \\
& \equiv & (\alpha\cos\Omega t +\beta\sin\Omega t)\hat{X}+(\beta\cos\Omega t-\alpha\sin
\Omega t)\hat{Y}+\gamma\hat{Z}.
\end{eqnarray}
The angle $\psi$ denotes the orientation of $\vec{B}$ in the lab frame; $\psi$ is the
angle of $\vec{B}$ north of the eastward direction. $\Omega$ is the Earth's
sidereal rotation frequency, and $t$ is local sidereal time. The orientation of
$\vec{B}$ depends on the three coefficients $\alpha=-\cos\psi$,
$\beta=-\sin\psi\cos\chi$ and $\gamma=\sin\psi\sin\chi$, with
$\alpha^{2}+\beta^{2}+\gamma^{2}=1$. $\alpha$, $\beta$, and $\gamma$ are determined by
the specific geometry of an individual collider experiment, as they parameterize the
direction of the local magnetic field in the laboratory.

The $\hat{B}_{X}$ and $\hat{B}_{Y}$ components of the field
are time dependent, but $\hat{B}_{Z}$
is not. According to (\ref{eq-coeffs}), the coefficients of the $c_{JK}c_{LM}$ in
$(\lambda_{1}-\lambda_{2})^{2}$ are polynomials in $\hat{B}_{X}$ and $\hat{B}_{Y}$
with degrees of up to four. This means that $(\lambda_{1}-\lambda_{2})^{2}$
can be expressed in terms of Fourier components as
\begin{eqnarray}
(\lambda_{1}-\lambda_{2})^{2} & = & {\cal A}_{0}+{\cal A}_{\Omega}\cos\Omega t+
{\cal B}_{\Omega}\sin\Omega t+{\cal A}_{2\Omega}\cos 2\Omega t+
{\cal B}_{2\Omega}\sin 2\Omega t \nonumber\\
\label{eq-fourier}
& & +{\cal A}_{3\Omega}\cos 3\Omega t+
{\cal B}_{3\Omega}\sin 3\Omega t+{\cal A}_{4\Omega}\cos 4\Omega t+
{\cal B}_{4\Omega}\sin 4\Omega t.
\end{eqnarray}
The Fourier coefficients ${\cal A}$ and ${\cal B}$, expressed in terms of the
$c_{JK}$, are rather complicated. However, they can be simplified substantially with
knowledge of which linear combinations of the coefficients this technique is
sensitive to.

In the Fourier series (\ref{eq-fourier}), the only coefficient that
depends on  $c_{ZZ}^{2}$ is ${\cal A}_{0}$. The reason is that the
the orientation of the magnetic field relative
to the $Z$-axis does not change as the Earth rotates. This means that there is no
sensitivity to the specific linear combination of coefficients
$c_{Q}=c_{XX}+c_{YY}-2c_{ZZ}$ (which parameterizes the effects of a preferred
direction parallel to the $Z$-axis) unless at least one other linear combination of
$c_{JK}$ coefficients is nonzero.
[The formula for ${\cal A}_{0}$ does contain a
term $\frac{1}{4}(\alpha^{2}+\beta^{2})c_{Q}^{2}$. However, while ${\cal A}_{0}$
represents a
genuine signal for anisotropy, the lack of any sidereal dependence tends to make this
signal indistinguishable from the effects of laboratory systematics.]

Most experiments that rely on the Earth's rotation do not exhibit any
sensitivity to $c_{Q}$ at ${\cal O}(c)$. However, since the observable $a/b$ is
a non-analytic function of the $c_{JK}$, the situation here differs a bit from the
usual one. There is no sensitivity to $c_{Q}$ alone. However, the sidereal
oscillations in $a/b$ do depend on the geometric means of $c_{Q}$ and other linear
combinations of the Lorentz violation coefficients.
So unless there is Lorentz violation of another sort, measurements of
the ${\cal A}$ and ${\cal B}$ coefficients cannot provide any constraints on
$c_{Q}$. In order to avoid ill conditioning in the problem of extracting the $c_{JK}$
coefficients from measured values of ${\cal A}$ and ${\cal B}$, it makes sense to
set $c_{Q}=0$. Doing so will simplify the expressions for the ${\cal A}$ and
${\cal B}$.

Moreover, since there is no sensitivity to the trace of $c_{JK}$ (which describes only
boost invariance violation), $c_{XX}$ and $c_{YY}$ can only enter in the combination
$c_{-}=c_{XX}-c_{YY}$. The forms of the coefficients of the time-dependent Fourier
components of are
\begin{eqnarray}
\label{eq-A1}
{\cal A}_{\Omega} & = &
\beta\gamma(\alpha^{2}+\beta^{2}-4)c_{-}c_{XZ}
+\alpha\gamma(\alpha^{2}+\beta^{2}-4)c_{-}c_{YZ} \nonumber\\
& & -2\alpha\gamma(\alpha^{2}+\beta^{2}-4)c_{XY}c_{XZ}
+2\beta\gamma(\alpha^{2}+\beta^{2}-4)c_{XY}c_{YZ} \nonumber\\
& & [+2\beta\gamma(\alpha^{2}+\beta^{2})c_{Q}c_{XZ}
-2\alpha\gamma(\alpha^{2}+\beta^{2})c_{Q}c_{YZ}] \\
{\cal B}_{\Omega} & = &
\alpha\gamma(\alpha^{2}+\beta^{2}-4)c_{-}c_{XZ}
-\beta\gamma(\alpha^{2}+\beta^{2}-4)c_{-}c_{YZ} \nonumber\\
& & +2\beta\gamma(\alpha^{2}+\beta^{2}-4)c_{XY}c_{XZ}
+2\alpha\gamma(\alpha^{2}+\beta^{2}-4)c_{XY}c_{YZ} \nonumber\\
& & [+2\alpha\gamma(\alpha^{2}+\beta^{2})c_{Q}c_{XZ}
+2\beta\gamma(\alpha^{2}+\beta^{2})c_{Q}c_{YZ}] \\
{\cal A}_{2\Omega} & = &
+2(\alpha^{4}-\beta^{4})c_{XZ}^{2}
+2(\beta^{4}-\alpha^{4})c_{YZ}^{2}
+8\alpha\beta(\alpha^{2}+\beta^{2})c_{XZ}c_{YZ} \nonumber\\
& & [-\frac{1}{2}(\alpha^{2}-\beta^{2})(\alpha^{2}+\beta^{2}-2)c_{-}c_{Q}
-2\alpha\beta(\alpha^{2}+\beta^{2}-2)c_{Q}c_{XY}] \\
{\cal B}_{2\Omega} & = &
-4\alpha\beta(\alpha^{2}+\beta^{2})c_{XZ}^{2}
+4\alpha\beta(\alpha^{2}+\beta^{2})c_{YZ}^{2}
+4(\alpha^{4}-\beta^{4})c_{XZ}c_{YZ} \nonumber\\
& & [+\alpha\beta(\alpha^{2}+\beta^{2}-2)c_{-}c_{Q}
-(\alpha^{2}-\beta^{2})(\alpha^{2}+\beta^{2}-2)c_{Q}c_{XY}] \\
{\cal A}_{3\Omega} & = &
\beta\gamma(\beta^{2}-3\alpha^{2})c_{-}c_{XZ}
+\alpha\gamma(\alpha^{2}-3\beta^{2})c_{-}c_{YZ} \nonumber\\
& & +2\alpha\gamma(\alpha^{2}-3\beta^{2})c_{XY}c_{XZ}
+2\beta\gamma(3\alpha^{2}-\beta^{2})c_{XY}c_{YZ} \\
{\cal B}_{3\Omega} & = & 
\alpha\gamma(3\beta^{2}-\alpha^{2})c_{-}c_{XZ}
+\beta\gamma(\beta^{2}-3\alpha^{2})c_{-}c_{YZ} \nonumber\\
&  & +2\beta\gamma(\beta^{2}-3\alpha^{2})c_{XY}c_{XZ}
+2\alpha\gamma(\alpha^{2}-3\beta^{2})c_{XY}c_{YZ} \\
{\cal A}_{4\Omega} & = &
\frac{1}{8}(\alpha^{4}-6\alpha^{2}\beta^{2}+\beta^{4})c_{-}^{2}
+2\alpha\beta(\alpha^{2}-\beta^{2})c_{-}c_{XY}
-\frac{1}{2}(\alpha^{4}-6\alpha^{2}\beta^{2}+\beta^{4})c_{XY}^{2} \\
\label{eq-B4}
{\cal B}_{4\Omega} & = &
\frac{1}{2}\alpha\beta(\beta^{2}-\alpha^{2})c_{-}^{2}
+\frac{1}{2}(\alpha^{4}-6\alpha^{2}\beta^{2}+\beta^{4})c_{-}c_{XY}
+2\alpha\beta(\alpha^{2}-\beta^{2})c_{XY}^{2}.
\end{eqnarray}
The expressions are given in
terms of $c_{-}$, $c_{XY}$, $c_{XZ}$, $c_{YZ}$, and $c_{Q}$. With the exception of
$c_{Q}$, only one of these linear combinations needs to be nonzero for there to be
an observable effect.  The terms that depend on $c_{Q}$ are provided for the sake of completeness only; these terms are printed in square brackets, so that they can be
dropped in practical calculations.

The expressions for the ${\cal A}$ and ${\cal B}$ Fourier coefficients are still
complicated, but they possess a number of readily understandable features. For
example,
if $\gamma=0$, indicating that $\vec{B}$ has no component in the $Z$-direction, the
Fourier coefficients for odd harmonics all vanish. In this case, the plane normal to
$\vec{B}$ is parallel to the $Z$-axis. As the laboratory rotates around that axis, the
plane returns to its original orientation after only half a sidereal day. However,
if $\gamma\neq0$, the normal plane is tilted relative to the $Z$-axis, and it takes
a full day for it to return to its original orientation.

If the Lorentz violation is purely in the $XY$-plane (only $c_{XX}=-c_{YY}$ and
$c_{XY}$ nonzero) only ${\cal A}_{4\Omega}$ and ${\cal B}_{4\Omega}$ are nonvanishing.
The major and minor axes of the orbit's projection interchange positions (in the lab frame) every quarter sidereal day; the ratio of the major to the minor thus
oscillates four times daily. Moreover, the amplitude of the oscillations in
$(\lambda_{1}-\lambda_{2})^{2}$ is proportional to $(c_{XX}-c_{YY})^{2}+4c_{XY}^{2}$,
which in this case is the square of the difference of the nonzero eigenvalues of the
matrix $c_{JK}$ itself.

To place constraints on the $c_{JK}$ coefficients requires a fit of the observed
$\epsilon^{4}=4(a/b-1)^{2}$ data for a given species. The fit should identify
oscillations at frequencies $\Omega$, $2\Omega$, $3\Omega$, and $4\Omega$.
Fermion and antifermion data can be combined and fitted together, because the
$c$ coefficients
distort particle and antiparticle orbits in the same fashion. The formulas
for the Fourier coefficients given in (\ref{eq-A1}--\ref{eq-B4}) can then be inverted
through a second fit, giving experimental values of the $c_{JK}$.

There are
eight potentially nonzero Fourier coefficients in (\ref{eq-fourier}),
making the
$c_{JK}$ values strongly overconstrained. Moreover, there are additional
observables in this system, which depend on the same underlying coefficients. If a
significant sidereal oscillation signal is seen in the ratio $a/b$, the potential Lorentz violation can be further checked by looking at the actual directions of the
major and minor axes, which will also oscillate in sidereal time.

\subsection{Eccentricities Dominated by Systematics}
\label{sec-systdom}

Section~\ref{sec-LVdom} discussed the behavior of the eccentricity in an idealized
situation, in which Lorentz violation is the dominant effect that deforms the
synchrotron orbits away from circularity. This is a potentially important regime,
and the
results are also interesting theoretically, since they introduce an ${\cal O}(c)$
observable that is a non-analytic function of the $c_{JK}$ parameters.
However, other effects may also lead to non-circular orbits, and a different
analysis is required if those other effects are more important than any Lorentz
violation.

It would be conceivable to have a systematic effect that made all orbits elliptical,
with the axis directions $\hat{a}$ and $\hat{b}$ fixed in the laboratory frame.
Alternatively, systematics might lead to a nonzero average eccentricity, but without
fixed $\hat{a}$ and $\hat{b}$. This might arise from changes in the detector
composition as a function of the distance from the beam axis. Different materials,
in which the orbiting particles have different energy loss rates, could lead to an
orbital deformation for which $\hat{a}$ or $\hat{b}$ is always oriented along the
radial direction.

If the orbital eccentricity is dominated by systematics, the effects of the
$c_{JK}$ may be treated as a perturbation. The primary Lorentz-violating effect
is still an extension of the major axis by a factor
$1-c^{\perp}_{jk}\hat{a}_{j}\hat{a}_{k}=1-c_{jk}\hat{a}_{j}\hat{a}_{k}$ and
contraction of the minor axis by
$1+c_{jk}\hat{b}_{j}\hat{b}_{k}$. (Since
$\hat{a}$ and $\hat{b}$ are then defined to lie in the plane normal to $\vec{B}$, the
use of $c^{\perp}$ instead of $c$ is unnecessary.)
The difference from the case discussed previously is that $\hat{a}$ and
$\hat{b}$ are not determined by $c$ itself but by the systematics. This
makes the eccentricity
\begin{equation}
\epsilon^{2}=\epsilon_{0}^{2}-2(1-\epsilon_{0}^{2})
(c_{jk}\hat{a}_{j}\hat{a}_{k}-c_{jk}\hat{b}_{j}\hat{b}_{k}),
\end{equation}
where $\epsilon_{0}$ is the eccentricity caused by systematics unrelated to
the Lorentz violation.

Since $\epsilon_{0}$ should not exhibit sidereal variations, it is still
possible to constrain the $c_{JK}$ coefficients by looking at the
time dependence of $\epsilon$. However, the data will need to be binned
according to the direction $\hat{a}$. If $\hat{a}$ is always radial and charged
particles are emitted from the interaction region in all directions, then the
average contribution that the $c_{JK}$ make to $\epsilon^{2}$ vanishes. For
some orbital orientations, $-2(1-\epsilon_{0}^{2})
(c_{jk}\hat{a}_{j}\hat{a}_{k}-c_{jk}\hat{b}_{j}\hat{b}_{k})$ will
increase $\epsilon^{2}$, but for an equal fraction of orbits, the
contribution to $\epsilon^{2}$ will be negative. However, if different $\hat{a}$
values are analyzed separately, this cancellation does not occur.


For orbits with major axis direction $\hat{a}$, the minor axis lies along
$\hat{b}=\hat{a}\times\hat{B}$. If
$\hat{a}(0)=\delta\hat{X}+\zeta\hat{Y}+\eta\hat{Z}$
at $t=0$, then the time-dependent axis directions are
\begin{eqnarray}
\hat{a}(t) & = & (\delta\cos\Omega t+\zeta\sin\Omega t)\hat{X}
+(\zeta\cos\Omega t-\delta\sin\Omega t)\hat{Y}+\eta\hat{Z} \\
\hat{b}(t) & = & [(\gamma\zeta-\beta\eta)\cos\Omega t+(\alpha\eta-\gamma\delta)
\sin\Omega t]\hat{X} \nonumber\\
& &
+\,[(\alpha\eta-\gamma\delta)\cos\Omega t-(\gamma\zeta-\beta\eta)\sin\Omega t]\hat{Y}
+(\beta\delta-\alpha\zeta)\hat{Z}.
\end{eqnarray}
This makes the observable quantity $\epsilon^{2}$ a second-order polynomial in
$\cos\Omega t$ and $\sin\Omega t$, expandable as
\begin{equation}
\epsilon^{2}=-2(1-\epsilon_{0}^{2})\left({\cal C}_{0}+{\cal C}_{\Omega}\cos\Omega t+
{\cal D}_{\Omega}\sin\Omega t+{\cal C}_{2\Omega}\cos 2\Omega t+
{\cal D}_{2\Omega}\sin 2\Omega t\right).
\end{equation}
The ${\cal C}$ and ${\cal D}$
Fourier coefficients that describe the sidereal variations
in this observable are
\begin{eqnarray}
\label{eq-C1}
{\cal C}_{\Omega} & = & 
2[\delta\eta-(\gamma\zeta-\beta\eta)(\beta\delta-\alpha\zeta)]c_{XZ}
+2[\zeta\eta-(\alpha\eta-\gamma\delta)(\beta\delta-\alpha\zeta)]c_{YZ} \\
{\cal D}_{\Omega}& = & 
2[\zeta\eta-(\alpha\eta-\gamma\delta)(\beta\delta-\alpha\zeta)]c_{XZ}
-2[\delta\eta-(\gamma\zeta-\beta\eta)(\beta\delta-\alpha\zeta)]c_{YZ} \\
{\cal C}_{2\Omega}& = & 
\frac{1}{2}[\delta^{2}-\zeta^{2}
-(\gamma\zeta-\beta\eta)^{2}+(\alpha\eta-\gamma\delta)^{2}]c_{-}
+2[\delta\zeta(\gamma\zeta-\beta\eta)(\alpha\eta-\gamma\delta)]c_{XY} \\
\label{eq-D2}
{\cal D}_{2\Omega} & = &
[\delta\zeta(\gamma\zeta-\beta\eta)(\alpha\eta-\gamma\delta)]c_{-}
-[\delta^{2}-\zeta^{2}
-(\gamma\zeta-\beta\eta)^{2}+(\alpha\eta-\gamma\delta)^{2}]c_{XY}.
\end{eqnarray}
The formulas depend on the directions
of $\hat{a}$ and $\vec{B}$ in the laboratory frame.

Generically, the sidereal variations in the eccentricity depend on the same four
$c_{JK}$ coefficients that were found to be observable
in section~\ref{sec-LVdom}:  $c_{-}$, $c_{XY}$,
$c_{XZ}$, and $c_{YZ}$. Since the effects of the Lorentz violation are, in this
case, perturbative---being added onto a larger systematic eccentricity---the
${\cal C}$ and ${\cal D}$ coefficients are simpler than the
${\cal A}$ and ${\cal B}$. The ${\cal C}$ and ${\cal D}$ depend only linearly on
the $c_{JK}$, rather than quadratically; this is related to the fact that the
natural observable is $\epsilon^{2}=2(a/b-1)$, rather than $\epsilon^{4}$.
The eight ${\cal A}$ and ${\cal B}$ Fourier coefficients provide redundant
measurements of the four relevant Lorentz violation parameters, but the
${\cal C}$ and ${\cal D}$ only yield four such measurements.
However, in both regimes there is
additional redundancy tied to measurement of the actual direction $\hat{a}$.
In the present analysis, that redundancy is explicit in the fact that the
$\epsilon^{2}$ data needs to be binned according to the laboratory value of
$\hat{a}$, so that each bin provides a separate measurement of the
eccentricity.

\section{Conclusion}
\label{sec-concl}

The Lorentz-violating background tensor
$c_{JK}$ contains five coefficients parameterizing anisotropy in the fermion
sector. One of these, $c_{Q}$, can be measured by looking at sidereal variations
in the orbital shape only if another linear combination is nonzero and dominates
any systematic effects. This leaves four
combinations of the $c_{JK}$ to be bounded independently.

For unstable particles,
almost any bounds that can be placed on these coefficients would be
of interest, since is quite difficult to constrain Lorentz violation for second- and
third-generation species. Measurements of the $c$ coefficients for the $t$ quark,
for example, are considered quite interesting, even though they may attain
only $10^{-2}$--$10^{-1}$ level sensitivity.
However, it would especially interesting if the laboratory constraints could
reach the same level of precision as the astrophysical photon survival bounds.
For $B$ mesons, this would mean placing bounds at approximately
the $10^{-7}$ level. Achieving
high sensitivities could be assisted by using the large statistics that are available
with collider data and by focusing on oscillations at a multiple of the Earth's sidereal frequency. However, for orbits with $\gtrsim 1$ m dimensions, reaching
$10^{-7}$ precision for the $c_{JK}$ would require
determining differences in average orbital dimensions at the sub-$\mu$m scale,
an extremely challenging proposition. With 100 $\mu$m precision,
two-sided constraints at the $10^{-4}$ level should be possible, and this would
still represent a major accomplishment for third-generation particles.

Although Lorentz violation is often associated with CPT violation, the $c$
coefficients are all even under C and CPT. For this reason, comparisons of the
orbital shapes for particles and antiparticles are of rather limited interest.
Renormalizable forms of Lorentz violation that do violate CPT tend to diminish in importance at high energies. The $d$ coefficients violate C (but not CPT) and do not
become less important at larger energies; however, the effects of the $d$ parameters
depend on both particle versus antiparticle identity and on helicity, making them
hard to measure without simultaneous measurements of particle polarizations.

While the orbital motion (\ref{eq-v}) was originally derived for fermionic particles,
there is an analogous set of coefficients for charged boson species. The orbits of
bosons are affected by Lorentz violation in essentially the same way as the orbits of
fermions. To apply the formulas in this paper to bosons, only the replacement
$c_{JK}\rightarrow\frac{1}{2}k_{JK}$ is required.

In summary, the Lorentz-violating $c$ coefficients break spatial isotropy.
Consequently, a particle with nonzero $c_{jk}$ coefficients will orbit elliptically
(instead of circularly) in a constant magnetic field. In an Earthbound detector,
the magnetic field direction will change as the Earth rotates, and so the
eccentricity $\epsilon$ of the orbital ellipse will be modulated at multiples of the
planet's sidereal rotation frequency. Measurements of these modulations can be turned
around to extract bounds on four of the five $c$ coefficients that describe
spin-independent spatial anisotropy. For most unstable fermion species, these would
represent the first two-sided constraints on these important SME parameters.

\section*{Acknowledgments}
The author is grateful to M. Purohit for helpful discussion.

\end{document}